\documentclass[ aps,%
floatfix,%
final,%
notitlepage,%
oneside,%
onecolumn,%
nobibnotes,%
nofootinbib,%
superscriptaddress,%
showpacs,%
]%
{revtex4-1}

\usepackage{graphicx}

\usepackage{amsmath}
\usepackage{hyperref}
\usepackage{booktabs}

\usepackage[english]{babel}

\newcommand{\Q}{\mathcal{Q}}

\newcommand{\M}{\mathcal{M}}
\newcommand{\A}{\mathcal{A}}
\newcommand{\I}{\mathcal{I}}

\newcommand{\C}{{\mathcal{C}}}
\newcommand{\JP}{{J/\psi}}
\newcommand{\fL}[1]{f_{#1}}
\newcommand{\phiL}[1]{\phi_{#1}}
\newcommand{\NRQCD}{\mathrm{NRQCD}}
\newcommand{\LC}{{\mathrm{LC}}}
\newcommand{\fN}[1]{F_{#1}}
\newcommand{\GeV}{\,\mathrm{GeV}}
\newcommand{\MeV}{\,\mathrm{MeV}}

\newcommand{\Br}{\mathrm{Br}}
\newcommand{\Ds}{{D_s^{(*)}}}
\newcommand{\fNCO}[1]{\tilde{F}_{{#1}}}

\usepackage{breqn}

\allowdisplaybreaks

\makeatletter
\let\cat@comma@active\@empty
\makeatother

\begin{document}
\title{Charmonia Production in $W\to (c\bar{c})\Ds$ Decays}

\author{A.~V.~Luchinsky}
\email{alexey.luchinsky@ihep.ru}
\affiliation{``Institute for High Energy Physics" NRC ``Kurchatov Institute'', 142281, Protvino, Russia}

\begin{abstract}
In the presented paper production  of charmonium state $\Q$ in exclusive $W\to\Q\Ds$ decays is analyzed in the framework of both leading order Nonrelativistic Quantum Chromodynamics (NRQCD) and light-cone expansion (LC) models. Analytical and numerical predictions for the branching fractions of these decays in both approaches are given. The typical value of the branching fractions is $\sim 10^{-11}$ and it turns out the LC results are about 4 times lager than NRQCD ones, so the effect of internal quark should be taken into account. Some estimates of color-octet contributions are presented and it is shown, that these contributions could be comparable with color-singlet results.
\end{abstract}
\maketitle

\section{Introduction}
\label{sec:introduction}

Heavy quarkonia mesons, i.e. particles that are build from heavy quark-antiquark pair are  very interesting states both from theoretical and experimental points of view. Because of the presence of two different mass scales the processes of their production and decays occur in two almost independent steps: production or decay of $(Q\bar{Q})$ pair and its hadronization into experimentally observed meson. Since the strong coupling constant $\alpha_s(m_Q)\ll 1$ the first step can be analyzed using perturbative QCD. Final hadronization, on the other hand, is essentially nonperturbative, so some other methods should be used.

One of such methods is Nonrelativistic Quantum Chromodynamics (NRQCD) \cite{Bodwin:1994jh}. In this approach the fact that the velocity of internal quark motion $v\sim\alpha_s(m_Q)$ is small in comparison with the speed of light and the probability of the considered process is written as a series over this small parameter. The hadronization probabilities are parametrized as NRQCD matrix elements, whose numerical values are determined, e.g. from solution of the potential models of analysis or the experimental data. Another interesting NRQCD feature is that production of color-octet (CO) components (when $Q\bar{Q}$-pair is in color-octet state and total color neutrality of the meson is guaranteed by the presence of additional gluons) can be considered. NRQCD approach was widely used for analysis of various processes and nice agreement with experimental results were achieved.

It should be noted, however, that in the case of charmonium meson production the NRQCD expansion parameter $v\sim\alpha_s(m_c)\sim 0.3$ is not really small, so the effect of internal quark motion should be taken into account. Another model for describing charmonium production at high energies is the so called light-cone (LC) expansion model \cite{Chernyak:1983ej}, when the amplitude of the reaction is written as a series over small chirality parameter $\sim m_c/E$, where $E$ is the typical energy scale of the considered reaction. This approach was also highly used in theoretical considerations of various reactions (see, e.g. \cite{Brambilla:2010cs, Braguta:2009df}) and often its predictions are more close to experimental data than NRQCD results. Usually the effect of internal quark motion leads to increase of theoretical predictions. For example, in the case of double charmonia production at B-factories Belle and BaBar there is about an order of magnitude difference between LC and NRQCD results and only LC solves the long standing contradiction between theory and experiment \cite{Abe:2002rb,Braaten:2002fi,Aubert:2005tj,Bondar:2004sv,Braguta:2005kr,Braguta:2006nf}

Recently a series of theoretical papers devoted to heavy quarkonia production in exclusive $W$-, $Z$-boson decays were published. For example in \cite{Wang:2013ywc,Grossmann:2015lea, Luchinsky:2017jab, Bodwin:2017pzj} charmonia $\Q$ production in radiative $Z$-boson decays $Z\to\Q\gamma$ was considered. In \cite{Likhoded:2017jmx} theoretical analysis of $Z\to\Q_1\Q_2$ decay was preformed. It is clear that in these processes the chirality expansion parameter $m_c/M_W\sim 2\times 10^{-2}$ is small, so LC framework can safely be used for their description. It was shown in mentioned above works that in both cases LC predictions are higher than NRQCD ones. In the presented paper we analyze charmonia production in exclusive $W\to\Q\Ds$ decays.

The rest of the paper is organized as follows. In the next section analytical results for the widths of the processes under consideration are given. Numerical predictions for the branching fractions both in color-singlet NRQCD and LC models are presented in section \ref{sec:numerical-results}. In section \ref{sec:color-octet} we give some estimates for CO contributions. The last section is reserved for conclusion.

\section{Analytical Results}
\label{sec:anslytical-results}

\begin{figure}
  \centering
  \includegraphics[width=0.6\textwidth]{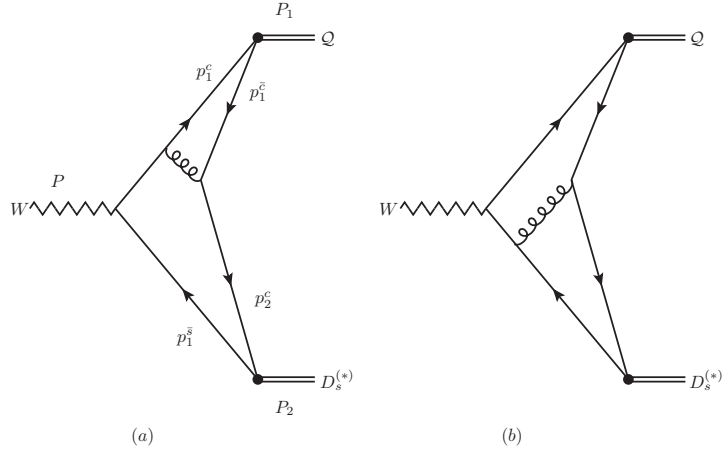}
  \caption{Typical Feynman diagrams for $W\to\Q\Ds$ decay in color-singlet approximation}
  \label{diagsCS}
\end{figure}

In our paper we consider charmonium meson $\Q$ production in exclusive $W$-boson decays
\begin{align}
  \label{eq:proc}
  W(P) &\to \Q(p_1) \Ds(p_2).
\end{align}
Typical Feynman diagrams describing this reaction are shown in Fig.~\ref{diagsCS}. In the current section we will restrict ourselves to color-singlet (CS) approximation, so only diagrams shown in Fig.~\ref{diagsCS} will contribute. Estimates for color-octet (CO) contributions will be given in section \ref{sec:color-octet}.

A widely used approach for description of heavy quarkonia  production is the Non-relativistic Quantum Chromodynamics (NRQCD) formalism \cite{Bodwin:1994jh}. In this model the amplitude of the process is written as a series over small quarks' relative velocity inside the meson.
At the leading order over this parameter internal quark motion is neglected completely, so quarks momenta are equal to
\begin{align}
  \label{eq:NRQCDmoms}
  p_{1}^{c} &= p_{1}^{\bar{c}}=\frac{m_c}{M_\Q}=\frac{P_1}{2},\qquad
   p_{2}^{c,\bar{s}}=\frac{m_{c,s}}{M_\Ds}P_2,
\end{align}
where $m_{c,s}$ are the masses of the corresponding quarks, $P_{1,2}$ are the momenta of charmonium and $\Ds$ mesons respectively, and $M_\Q=2m_c$, $M_\Ds=m_c+m_s$ are their masses. The projection on physical states is calculated using the technique described, e.g. in paper \cite{Braaten:2002fi}. It turns out that after straightforward (although rather cumbersome) calculations this approach leads to surprisingly simple expressions for the widths of the considered processes:
\begin{align}
  \label{eq:NRQCDwidth}
  \Gamma_\NRQCD\left(W\to\Q\Ds\right) &= \frac{16\pi\alpha_s^2 \lambda g_W^2}{243}
         \left(\frac{m_c+m_s}{m_c}\right)^2
         \frac{\fN{\Q}^2 \fN{\Ds}^2}{M_W^3} C_{\Q\Ds},
\end{align}
where $\alpha_s=\alpha_s(M_W)$ is a strong coupling constant,
\begin{align}
  \label{eq:gW}
  g_W &=\frac{e V_{cs}}{2\sqrt{2}\sin\theta_W}
\end{align}
is the $W\to cs$ vertex coupling constant,
\begin{align}
  \label{eq:lambda}
  \lambda &= \sqrt{1-\left(\frac{M_Q-M_\Ds}{M_W}\right)^2} \sqrt{1-\left(\frac{M_Q+M_\Ds}{M_W}\right)^2}
\end{align}
is final meson's velocity in $W$ rest frame, and  $F_{\Q,D_s^{(*)}}$ are longitudinal constants of the final mesons defined as
\begin{align}
  \label{eq:f_NRQCD_S}
 \fN{\eta_c} &= \fN{\JP} = \sqrt{\frac{\langle O_1\rangle_{\JP}}{m_c}}, \\
  \label{eq:f_NRQCD_P}
 \fN{h_c} &= \sqrt{3} \fN{\chi_{c0}} = \frac{1}{\sqrt{2}}\fN{\chi_{c1}} = \sqrt{\frac{3}{2}} \fN{\chi_{c2}} = \sqrt{\frac{\langle O_1\rangle_{h_C}}{m_c^3}},   
\end{align}
where $\langle O_1\rangle_{J/\psi,h_c}$ are defined in \cite{Braaten:2002fi} NRQCD matrix elements for $S$- and $P$-wave charmonium mesons. As for  dimensionless coefficients $C_{\Q D_s^{(*)}}$, expressions for them are presented in the Appendix. It should be mentioned, however, that in massless quark limit $m_{s,c}\ll m_W$ we have $C_{\Q D_s^{(*)}}=1$.

An alternative way to calculate the widths of the considered processes is so called light-cone (LC) expansions formalism \cite{Chernyak:1983ej}. In the framework of this method the amplitude of the reaction is written as a series over small chirality parameter $\sim m_q/M_W$. According to LC selection rules the total helicity of the hadronic states should be conserved. There are only two hadrons in the reaction, so from this rule it follows that
\begin{align}
  \label{eq:hel1}
  \lambda_1+\lambda_2 &=0
\end{align}
, where $\lambda_{1,2}$ are the helicities of final mesons. Orbital momentum conservation, in the other hand, requires
\begin{align}
  \label{eq:hel2}
  -1 &\le \lambda_W=\lambda_1-\lambda_2\le 1.
\end{align}
It is clear that only $\lambda_1=\lambda_2=0$ satisfy both of these restrictions, so only production of longitudinally polarized mesons is allowed at the leading twist approximation. Thus, in the framework of LC formalism the amplitude of $W\to\Q\Ds$ decay equals to
\begin{align}
  \label{eq:LCmatr}
  \M(W\to\Q\Ds) &\sim \fL{\Q}\fL{\Ds} \int_{-1}^1 d\xi_1 d\xi_2 \phiL{\Q}(\xi_1)\phiL{\Ds}(\xi_2)\A,
\end{align}
where $\xi_{1,2}=2x_{1,2}-1$ with $x_{1,2}$ being the momentum fractions of $c$-quarks inside $\Q$ and $\Ds$ mesons, $\phiL{\Q,\Ds}(\xi_{1,2}$ are light-cone distribution functions, $\fL{\Q,\Ds}$ are longitudinal mesonic constants, and the amplitude $\A$ can be calculated using perturbation QCD on the basis of presented in Fig.~\ref{diagsCS} diagrams. According to \cite{Chernyak:1983ej} mesonic constants and distribution amplitudes are defined as
\begin{align}
  \label{eq:DA_even}
  \langle\Q_L(p)
  \left|\bar{c}^i_\alpha(z)[z,-z]c^j_\beta(-z)\right|0\rangle
  &=
    \left(\hat{p}\right)_{\alpha\beta}\frac{\fL{\Q}}{4}
    \frac{\delta^{ij}}{3}
    \int\limits_{-1}^1 \phiL{\Q}(\xi)d\xi
\end{align}
for $\sigma$-even states $\Q=\JP$, $\chi_{0,2},\Ds$, and
\begin{align}
  \label{eq:DA_odd}
  \langle\Q_L(p)
  \left|\bar{c}^i_\alpha(z)[z,-z]c^j_\beta(-z)\right|0\rangle
  &=
    \left(\hat{p}\gamma_5\right)_{\alpha\beta}\frac{\fL{\Q}}{4}
    \frac{\delta^{ij}}{3}
    \int\limits_{-1}^1 \phiL{\Q}(\xi)d\xi
\end{align}
for $\sigma$-odd states $\Q=\eta_c$, $\chi_{c1}$, and $h_c$. These definitions for LC constants is consistent with NRQCD mesonic constants \eqref{eq:f_NRQCD_S}, \eqref{eq:f_NRQCD_P}. In the above expressions $\alpha,\beta$ and $i,j$ are spinor and colour indices of quark and antiquark respectively. The normalization condition for the distribution amplitudes is
\begin{align}
  \label{eq:norm_xi_even}
  \int_{-1}^1\phiL{\Q}(\xi)d\xi &=1,\qquad   \int_{-1}^1 \xi\phiL{\Q}(\xi)d\xi=1 
\end{align}
for $\xi$-even  ($\Q=\eta_c,\JP, \chi_{c1}, \Ds$)  and $\xi$-odd ($\Q=\chi_{c0,2}, h_c$) states. In $\delta$ approximation, when internal quark motion is neglected, the distribution amplitudes of charmonia states take the form
\begin{align}
  \label{eq:deltaCC}
  \phiL{\eta_c,\JP,\chi_{c1}}(\xi) &= \delta(\xi),\qquad \phiL{\chi_{c0,2},h_c}(\xi)=-\delta'(\xi),
\end{align}
while for $\Ds$ mesons we have
\begin{align}
  \label{eq:deltaDs}
  \phiL{\Ds} &= \delta\left(\xi-\frac{m_c-m_s}{m_c+m_s}\right).
\end{align}

The light-cone amplitude corresponding to presented in Fig.~\ref{diagsCS} diagrams is equal to
\begin{align}
  \label{eq:matrLC}
  \M(W\to \Q\Ds) &= \frac{16\pi\alpha_s g_W}{9} \frac{\fL{\Q}\fL{\Ds}}{M_W^2} \I_{\Q\Ds} (P_1-P_2)_\mu\epsilon_W^\mu,
\end{align}
where $\epsilon_W^\mu$ is the polarization vector of the initial $W$-boson and
\begin{align}
  \label{eq:I}
  \I_{\Q\Ds} &= 2\int_{-1}^2 d\xi_1 d\xi_2 \frac{\phiL{\Q}(\xi_1)\phiL{\Ds}(\xi_2)}{(1-\xi_1)(1+\xi_2)}.
\end{align}
The corresponding width is equal to
\begin{align}
  \label{eq:widthLC}
  \Gamma_\LC\left(W\to\Q\Ds\right) &= \frac{16\pi\alpha_s^2g_W^2}{243}\frac{\fL{\Q}^2\fL{\Ds}^2}{M_W^2} \I_{\Q\Ds}^2.
\end{align}
It is easy to check that in $\delta$-approximation \eqref{eq:deltaCC}, \eqref{eq:deltaDs} NRQCD result \eqref{eq:NRQCDwidth} is restored.

\section{Numerical Results}
\label{sec:numerical-results}

Let us first consider NRQCD predictions for the widths of $W\to\Q\Ds$ decays. Numerical values of final meson's masses were taken from PDG tables \cite{Patrignani:2016xqp} and quarks' masses were chosen to be equal to
\begin{align}
  \label{eq:qMasses}
  m_c &= \frac{M_{\Q}}{2},\qquad m_s=M_{\Ds}-m_c.
\end{align}
The mesonic constants, entering relation \eqref{eq:NRQCDwidth} can be related to  matrix elements $\left<O_1\right>_{J/\psi,h_c}$ using relations \eqref{eq:f_NRQCD_S}, \eqref{eq:f_NRQCD_P}, where
\cite{Braaten:2002fi}
\begin{align}
  \label{eq:NRQCD_O_num}
  \langle O_1\rangle_{\JP} &= 0.22\,\GeV^3,\qquad \langle O_1\rangle_{h_c} = 0.033\,\GeV^3.
\end{align}
These values correspond to
\begin{align}
  \label{eq:fNRQCD_num}
 \fN{\eta_c} &= \fN{\JP}=0.38\,\GeV,\\
  \fN{\chi_{c0}} &= 0.057\,\GeV, \fN{\chi_{c1}}=0.14,\GeV, \\
  \fN{\chi_{c2}} &=0.081\,\GeV, \fN{h_c}=0.099\,\GeV.
\end{align}
The strong coupling constant $\alpha_s(\mu^2)$ is parametrized as
 \begin{align}
   \label{eq:alphas}
   \alpha_s(\mu^2) &= \frac{4\pi}{b_0\ln(\mu^2/\Lambda_{\mathrm{QCD}^2)}}, \quad b_0=11-\frac{2}{3}n_f,
 \end{align}
 where $\Lambda_{\mathrm{QCD}}\approx 0.2\,\GeV$ and $n_f=5$ is the number of active flavors. At the scale $\mu^2=M_W^2$ it corresponds to $\alpha_s(M_W^2)=0.14$. With presented above values of the parameters it is easy to obtain branching fractions presented in the second columns of tables \ref{tab:brDs}, \ref{tab:brDs_star}.

\begin{table}[h]
  \centering
  \begin{tabular}{lllll}
    \hline
    $Q$ & $\Br_\NRQCD, 10^{-12}$ & $\Br_\LC^\delta, 10^{-12}$ & $\Br_\LC, 10^{-12}$ & $\Br_\LC/\Br_\LC^\delta$\\
    \hline
    $\eta_c$ & $2.28$ & $3.\pm 0.4 $ & $13.1\pm 2._{-0.84}^{+2.7}$ & $4.37_{-0.3}^{0.9}$ \\ 
$J/\psi$ & $2.1$ & $4.12\pm 0.4 $ & $18.\pm 2._{-1.1}^{+3.7}$ & $4.37_{-0.3}^{0.9}$ \\ 
$h_c$ & $0.112$ & $0.906\pm 0.3 $ & $2.13\pm 0.8_{-0.24}^{+0.55}$ & $2.35_{-0.3}^{0.6}$ \\ 
$\chi_{c0}$ & $0.0387$ & $0.302\pm 0.1 $ & $0.71\pm 0.3_{-0.079}^{+0.18}$ & $2.35_{-0.3}^{0.6}$ \\ 
$\chi_{c1}$ & $0.226$ & $1.81\pm 0.7 $ & $7.83\pm 3._{-0.53}^{+1.6}$ & $4.32_{-0.3}^{0.9}$ \\ 
$\chi_{c2}$ & $0.0731$ & $0.604\pm 0.2 $ & $1.42\pm 0.5_{-0.16}^{+0.37}$ & $2.35_{-0.3}^{0.6}$ \\ 
   \hline
  \end{tabular}
  \caption{Branching fractions of $W\to Q D_s$ decays. In second, third and fourth columns results of NRQCD formalism, LC approach in $\delta$-approximation \eqref{eq:deltaCC}, \eqref{eq:deltaDs}, and LC results with \eqref{eq:phiLCS}, \eqref{eq:phiLCPeven}, \eqref{eq:phiLCPodd}, \eqref{eq:phi_appr} distribution amplitudes are given. In the last column of the table the effect of internal quark motion is shown. The uncertainties in $\Br_\LC^\delta$ predictions and first errors in $\Br_\LC$ predictions  are caused by mesonic constants uncertainties \eqref{eq:fLC_num}, second errors in fourth and the error in the last column are caused by the variation of distribution amplitudes' parameters \eqref{eq:beta}, \eqref{eq:as}.
}
  \label{tab:brDs}
\end{table}

\begin{table}[h]
  \centering
  \begin{tabular}{lllll}
    \hline
    $Q$ & $\Br_\NRQCD, 10^{-12}$ & $\Br_\LC^\delta, 10^{-12}$ & $\Br_\LC, 10^{-12}$ & $\Br_\LC/\Br_\LC^\delta$\\
    \hline
    $\eta_c$ & $3.18$ & $3.38\pm 0.5 $ & $14.8\pm 2._{-0.95}^{+3.}$ & $4.37_{-0.3}^{0.9}$ \\ 
$J/\psi$ & $2.97$ & $4.64\pm 0.5 $ & $20.3\pm 2._{-1.3}^{+4.1}$ & $4.37_{-0.3}^{0.9}$ \\ 
$h_c$ & $0.153$ & $1.02\pm 0.4 $ & $2.4\pm 0.9_{-0.27}^{+0.62}$ & $2.35_{-0.3}^{0.6}$ \\ 
$\chi_{c0}$ & $0.0664$ & $0.341\pm 0.1 $ & $0.8\pm 0.3_{-0.089}^{+0.21}$ & $2.35_{-0.3}^{0.6}$ \\ 
$\chi_{c1}$ & $0.311$ & $2.04\pm 0.8 $ & $8.83\pm 3._{-0.6}^{+1.8}$ & $4.32_{-0.3}^{0.9}$ \\ 
$\chi_{c2}$ & $0.102$ & $0.681\pm 0.3 $ & $1.6\pm 0.6_{-0.18}^{+0.41}$ & $2.35_{-0.3}^{0.6}$ \\ 
   \hline
  \end{tabular}
  \caption{Branching fractions of  $W\to Q D_s^*$ decays. Notations are the same as in table \ref{tab:brDs}}
  \label{tab:brDs_star}
\end{table}

In order to calculate  LC predictions (see relations \eqref{eq:widthLC}, \eqref{eq:I}) numerical values of mesonic constants $\fL{\Q}$ and parametrization for distribution amplitudes $\phiL{\Q}(\xi)$ is required. It should be noted that both $\fL{}$ and $\phiL{}(\xi)$ actually depend on the renormalization scale $\mu$ (see \cite{Gribov:1972rt,Lipatov:1974qm,Altarelli:1977zs,Braun:2003rp}). According to \cite{Braguta:2006wr,Braguta:2007fh,Braguta:2008qe} the following values of the constants will be used:
\begin{align}
  \label{eq:fLC_num}
  \fL{\eta_c}(m_c) &=(0.35\pm0.02)\GeV, \quad \fL{\JP}(m_c)=(0.41\pm0.02)\GeV,\quad
  \fL{\chi_{c0}}(m_c) = (0.11\pm0.02)\GeV,\\
 \fL{\chi_{c1}}(m_c) &= (0.27\pm0.05)\GeV,\quad
  \fL{\chi_{c2}}(m_c)  = (0.16\pm0.03)\GeV, \quad \fL{h_c}(m_c) = (0.19\pm0.03)\GeV.
\end{align}
Note, the value of $c$ quark in LC model differs from phenomenological choice $M_{\JP}/2$ and is equal to $m_c=1.2\GeV$. As for $\Ds$ mesons, in the following we will use \cite{RandleConde:2011xq, Donald:2013sra}
\begin{align}
  \label{eq:fLCDs_num}
  \fL{D_s}(m_c) &= 258\,\MeV,\qquad \fL{D_s^*}(m_c) = 274\,\MeV.
\end{align}

\begin{figure}
  \centering
  \includegraphics[width=0.9\textwidth]{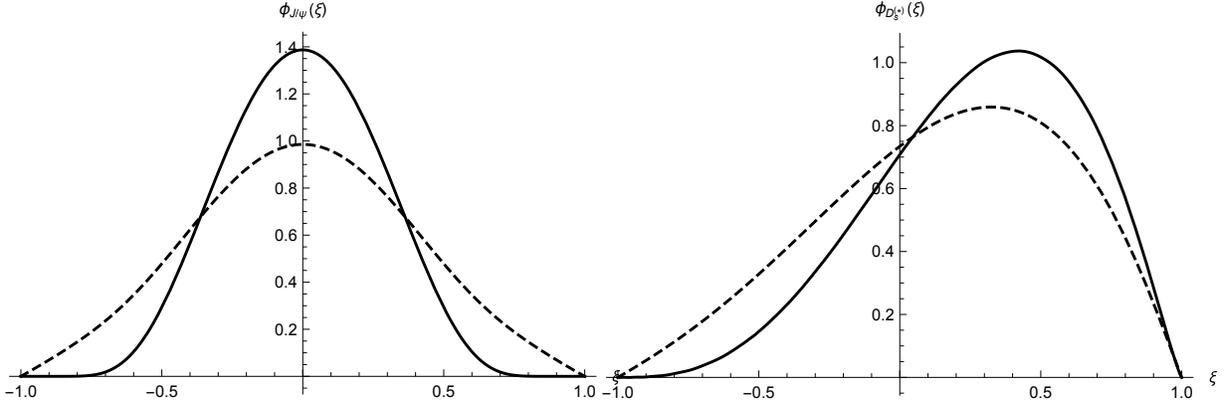}
  \caption{Distribution amplitudes for $S$-wave mesons $\eta_c$, $\JP$ (left figure) and $\Ds$ (right figure). Solid and dashed lines correspond to $\mu=m_c$ and $\mu=M_W$ respectively. }
  \label{fig:DAS}
\end{figure}

\begin{figure}
  \centering
  \includegraphics[width=0.9\textwidth]{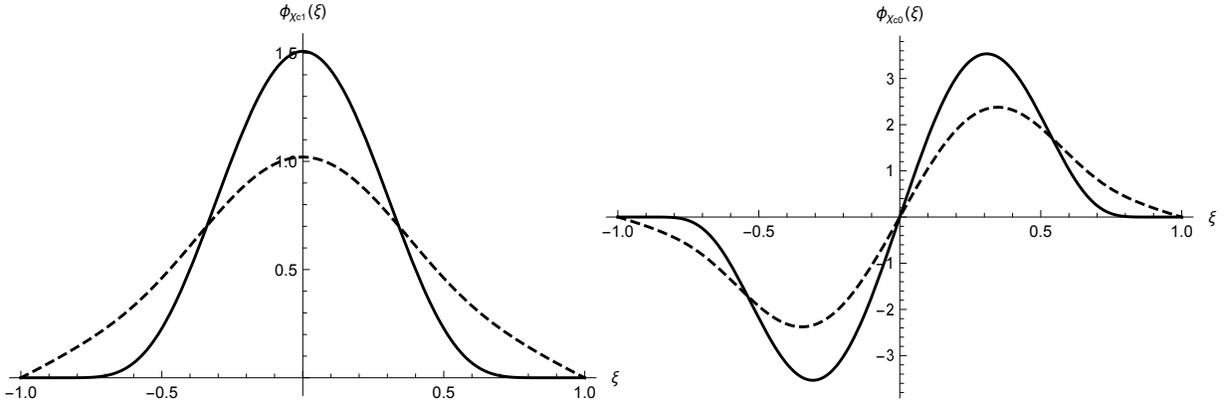}
  \caption{Distributions amplitudes for $P$-wave mesons $\chi_{c1}$ (left figure) and $\chi_{c0,2}$, $h_c$ (right figure). Notations are the same as in Fig.\ref{fig:DAS}.}
  \label{fig:DAP}
\end{figure}

The distribution amplitude of charmonium mesons can be written in the form \cite{Braguta:2006wr, Braguta:2007fh, Braguta:2008qe}
\begin{align}
  \label{eq:phiLCS}
  \phiL{\JP,\eta_c}(\xi,\mu_0) &= c(\beta_S) (1-\xi^2) \exp\left(-\frac{\beta_S}{1-\xi^2}\right),\\
  \label{eq:phiLCPeven}
  \phiL{\chi_{c0},\chi_{c2},h_c}(\xi,\mu_0) &= c_1(\beta_P) \xi(1-\xi^2) \exp\left(-\frac{\beta_P}{1-\xi^2}\right),\\
  \label{eq:phiLCPodd}
    \phiL{\chi_{c1}}(\xi,\mu_0) &= -c_2(\beta_P)\int_{-1}^\xi \phiL{h_c}(\xi,\mu_0),
\end{align}
where $c(\beta_S)$, $c_{1,2}(\beta_P)$ are normalization constants \eqref{eq:norm_xi_even} and wave function parameters are equal to
\begin{align}
  \label{eq:beta}
  \beta_S &=3.8\pm0.7,\qquad  \beta_P =3.4^{+1.5}_{-0.9}.
\end{align}
Distribution amplitudes of  $\Ds$ meson at $\mu=m_c$ will be parametrized as
\begin{align}
  \label{eq:phi_appr}
  \phiL{}(\xi) &\sim (1-\xi)^{a_c} (1+\xi)^{a_s},
\end{align}
where, according to \cite{Kartvelishvili:1977pi, Gershtein:2006ng}, $a_c=3.1$ and
\begin{align}
  \label{eq:as}
  1&\le a_s<2
\end{align}
For mean $\xi$ value NRQCD limit
\begin{align}
  \label{eq:xi0}
  \langle \xi \rangle &= \frac{m_c-m_s}{m_c+m_s}
\end{align}
is observed at $a_s\approx1.2$. In Figures \ref{fig:DAS}, \ref{fig:DAP} we show the distributions amplitudes at different scales. From these figures it is clear that with the increase of the scale effective width of the distribution amplitude also increases.

In  tables \ref{tab:brDs}, \ref{tab:brDs_star} we show LC predictions for branching fractions of the considered decays in $\delta$-approximation (third columns) and using real distribution amplitudes \eqref{eq:phiLCS}, \eqref{eq:phiLCPeven}, \eqref{eq:phiLCPodd}, \eqref{eq:phi_appr} with mentioned above values of the parameters $\beta_{S,P}$ and $a_s$ (fourth columns). In the fifth columns of the tables the effect of internal quark motion is shown. It can be easily seen that as a result of this effect the branching fractions of the decays increase significantly.

\section{Color-Octet Contributions}
\label{sec:color-octet}

As it was shown in the previous section, in spite of the increase caused by internal quark motion the branching fractions of the considered decays are small. This is caused mainly by the large value of $W$-boson mass, that enters in shown in Fig.~\ref{diagsCS} gluon propagators. It is clear, on the other hand, that in the case of color-octet (CO) state production the situation is completely different. In the current section we will give rough estimates for CO contributions.

\begin{figure}
  \centering
  \includegraphics[width=0.9\textwidth]{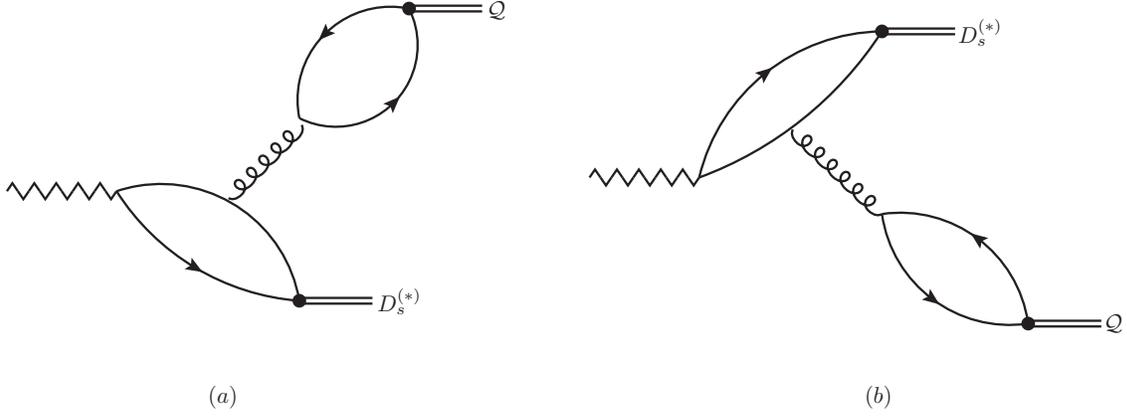}
  \caption{Feynman diagrams for $W\to\Q\Ds$ decays in CO approximation}
  \label{diagsCO}
\end{figure}

In addition to shown in Fig.~\ref{diagsCS} Feynman diagrams shown in Fig.~\ref{diagsCO} diagrams also contribute to the process under consideration in color-octet approximation. In order to calculate the corresponding decay width it is convenient simply to change in the projection operator $\fN{\Q}$ mesonic constant to color-octet parameter $\fNCO{\Q}$ and color identity matrix $\delta^{ij}/\sqrt{N_c}$ to the corresponding Gell-mann matrix $\sqrt{2}(T^a)_{ij}$, where $N_c=3$ is the number of colors \cite{Cho:1995ce}. With these substitutions the width of $W\to J/\psi D_s$ decay in CO approximation takes the form
\begin{align}
  \label{eq:gammaNRQCDco}
  \Gamma_\NRQCD^{CO} &\approx \frac{\pi\alpha_s^2 g_W^2}{54}\frac{(m_c^2+m_c^2)(m_c+m_s)^2}{m_c^4m_s^2} \frac{\fNCO{J/\psi}^2 \fNCO{D_s}^2}{M_W} \C_{J/\psi D_s}^{CO},
\end{align}
where
\begin{align}
  \C_{J/\psi D_s}^{CO} &=
                         \Bigg[m_s^2 (3717 m_c^{16}+21990 m_c^{15} m_s+m_c^{14} (39548 m_s^2+4435 M_W^2)+m_c^{13} (39076 m_s M_W^2-109006 m_s^3)+
\nonumber\\&                          
m_c^{12} (-135880 m_s^4+8281 m_s^2 M_W^2+3153 M_W^4)+2 m_c^{11} (64307 m_s^5-41908 m_s^3 M_W^2-13555 m_s M_W^4)+
\nonumber\\&                          
m_c^{10} (195276 m_s^6+1123 m_s^4 M_W^2-53870 m_s^2 M_W^4-5309 M_W^6)-2 m_c^9 (14783 m_s^7+514 m_s^5 M_W^2-
\nonumber\\&                          
51765 m_s^3 M_W^4+5028 m_s M_W^6)+m_c^8 (-125238 m_s^8+16465 m_s^6 M_W^2+38815 m_s^4 M_W^4+32583 m_s^2 M_W^6-
\nonumber\\&                          
449 M_W^8)-2 m_c^7 (19247 m_s^9-12328 m_s^7 M_W^2+6678 m_s^5 M_W^4+7280 m_s^3 M_W^6-2637 m_s M_W^8)+
\nonumber\\&                          
m_c^6 (28788 m_s^{10}-623 m_s^8 M_W^2-37764 m_s^6 M_W^4+7686 m_s^4 M_W^6-6664 m_s^2 M_W^8+705 M_W^{10})+
\nonumber\\&                          
2 m_c^5 (9635 m_s^{11}-5426 m_s^9 M_W^2-4030 m_s^7 M_W^4-3432 m_s^5 M_W^6+3555 m_s^3 M_W^8-302 m_s M_W^{10})+
\nonumber\\&                          
m_c^4 (m_s^2-M_W^2)^2 (2288 m_s^8+2659 m_s^6 M_W^2+1653 m_s^4 M_W^4+1517 m_s^2 M_W^6-117 M_W^8)-
\nonumber\\&                          
2 m_c^3 m_s (m_s^2-M_W^2)^3 (503 m_s^6-167 m_s^4 M_W^2-259 m_s^2 M_W^4+51 M_W^6)-m_c^2 (m_s^2-M_W^2)^4 (316 m_s^6+311 m_s^4 M_W^2+
\nonumber\\&                          
98 m_s^2 M_W^4-9 M_W^6)+6 m_c m_s^3 (m_s^2-5 M_W^2) (m_s^2-M_W^2)^5+
\nonumber\\&                          
9 m_s^2 (m_s^2-M_W^2)^6 (m_s^2+M_W^2))\Bigg]/\Bigg[9 M_W^2 (m_c^2+m_s^2) (m_c-m_s+M_W)^4 (-m_c+m_s+M_W)^4 (-4 m_c^3+
\nonumber\\&                          
  m_c^2 m_s+2 m_c m_s^2+m_s^3-m_s M_W^2)^2\Bigg]
\end{align}
is equal to 1 in large $M_W$ limit. It can easily be seen, that the decrease of this decay width with $M_W$ is much slower than in CS case:
\begin{align}
  \label{eq:COrat}
  \frac{\Gamma_\NRQCD^{CO}}{\Gamma_\NRQCD} &\sim \frac{M_W^2}{m_c^2}.
\end{align}
This behavior is explained by the fact that gluon virtuality in color-octet diagrams is $M_{J/\psi}^2$ instead of typical oder $M_W^2$ in the case of color-singlet mechanism.
Numerical calculations show that  for optimistic assumptions for color-octet constants $\fNCO{}\sim 10^{-1}\fN{}$ the contribution of CO mechanism is comparable with CS one.

\section{Conclusion}
\label{sec:conclusion}

In the presented article production of charmonium $\Q$ in exclusive $W\to \Q\Ds$ decays is analyzed using both Non-relativistic Quantum Chromodynamics (NRQCD) and light-cone expansion (LC) approaches.

Presented in the paper theoretical NRQCD predictions show, that the branching fractions of the considered decays are pretty small, although about an order of magnitude higher than obtained in the previous work \cite{Likhoded:2017jmx} branching fractions of double charmonium production in exclusive $Z$-boson decays. The effect of internal quarks' motion, analyzed using LC formalism increases the branching fractions significantly, but they still remains small. The reason for this fact is twofold:
\begin{itemize}
\item In contrast to $Z\to 2\Q$ decay two $S$-wave mesons can be produced at the leading twist approximation, so no chirality suppression factors occur,
\item Production of the lighter system $\Q\Ds$ instead of $\Q_1\Q_2$ one make the probability of the process larger,
\item The widths of the considered decays are, nevertheless, suppressed my large $W$-boson mass ($\sim 1/M_W^3$), so the branching fractions are small.
\end{itemize}
The last point can be bypassed if production of color-octet (CO) states is considered. In the last section of the article we give rough estimates for CO contributions and show that $M_W$ suppression of the resulting width is not so strong and the behavior $\Gamma\sim 1/M_W$ is observed. In the case of $J/\psi D_s$ pair production,
our calculations show, that with reasonable assumtions on the value of color-octet matrix elements the resulting widths are comparable with color-singlet ones.

In our future work we plan to analyze production of other states (e.g. excited charmonia and $P$-wave charmonium mesons CO states) in more details.

All calculations in the article were performed with the help of FeynCalc Mathematica package \cite{Mertig:1990an,Shtabovenko:2016sxi}. The author would like to thank A.~K.~Likhoded for fruitful discussions.


%

\appendix

\section{NRQCD Widths}
\label{sec:nrqcd-cross-sections}

Below we give explicit expressions for $C_{\Q\Ds}$ coefficients defined in equation (\ref{eq:NRQCDwidth}). It is convenient to introduce dimensionless variables
\begin{align}
  r_c &= \frac{m_c}{M_W},\quad r_s=\frac{m_s}{M_W},\qquad X=1-(r_c-r_s)^2.
\end{align}
With these notations we have
\begin{dgroup*}
\begin{dmath*}
X^3 \C_{\eta_c D_s} =1+r_c^2+6 r_c r_s+r_s^2-65 r_c^4-108 r_c^3 r_s-54 r_c^2 r_s^2-12 r_c r_s^3-r_s^4-\left(15 r_c^3+23 r_c^2 r_s+9 r_c r_s^2+r_s^3\right)^2,
\end{dmath*}
\begin{dmath*}
  X^4 \C_{\eta_c D_s^*} =1+4 \left(-r_c^2+4 r_c r_s+r_s^2\right)-95 r_c^4-348 r_c^3 r_s-226 r_c^2 r_s^2-76 r_c r_s^3-7 r_s^4+2 \left(247 r_c^6+774 r_c^5 r_s+721 r_c^4 r_s^2+436 r_c^3 r_s^3+121 r_c^2 r_s^4+6 r_c r_s^5-r_s^6\right)+
  4 \left(-3 r_c^4-10 r_c^3 r_s+6 r_c^2 r_s^2+6 r_c r_s^3+r_s^4\right)^2,
\end{dmath*}
\begin{dmath*}
X^4 \C_{J/\psi D_s} =1-2 \left(7 r_c^2+6 r_c r_s+3 r_s^2\right)+53 r_c^4+68 r_c^3 r_s+86 r_c^2 r_s^2+52 r_c r_s^3+13 r_s^4-4 \left(7 r_c^6-78 r_c^5 r_s-83 r_c^4 r_s^2-20 r_c^3 r_s^3+25 r_c^2 r_s^4+18 r_c r_s^5+3 r_s^6\right)+4 \left(3 r_c^4-8 r_c^3 r_s+4 r_c r_s^3+r_s^4\right)^2,
\end{dmath*}
\begin{dmath*}
X^4 \C_{J/\psi D_s^*} =1+10 (r_c+r_s)^2-2 \left(87 r_c^4+148 r_c^3 r_s+118 r_c^2 r_s^2+44 r_c r_s^3+11 r_s^4\right)+2 \left(53 r_c^6-34 r_c^5 r_s-21 r_c^4 r_s^2+36 r_c^3 r_s^3+59 r_c^2 r_s^4+30 r_c r_s^5+5 r_s^6\right)+\left(17 r_c^2+2 r_c r_s+r_s^2\right) \left(-3 r_c^3-r_c^2 r_s+3 r_c r_s^2+r_s^3\right)^2,
\end{dmath*}
\begin{dmath*}
X^4 \C_{h_c D_s} =1+2 \left(7 r_c^2+8 r_c r_s+r_s^2\right)-83 r_c^4-92 r_c^3 r_s-86 r_c^2 r_s^2-52 r_c r_s^3-7 r_s^4-4 \left(100 r_c^6+235 r_c^5 r_s+227 r_c^4 r_s^2+94 r_c^3 r_s^3-6 r_c^2 r_s^4-9 r_c r_s^5-r_s^6\right)+4 r_c^2 \left(33 r_c^3+41 r_c^2 r_s+19 r_c r_s^2+3 r_s^3\right)^2,
\end{dmath*}
\begin{dmath*}
X^5 \C_{h_c D_s^*} =1-19 r_c^2-6 r_c r_s+r_s^2+4 \left(26 r_c^4-12 r_c^3 r_s+r_c^2 r_s^2+10 r_c r_s^3-r_s^4\right)+4 \left(92 r_c^6+602 r_c^5 r_s+415 r_c^4 r_s^2+168 r_c^3 r_s^3+30 r_c^2 r_s^4-26 r_c r_s^5-r_s^6\right)-3649 r_c^8-15624 r_c^7 r_s-20880 r_c^6 r_s^2-20928 r_c^5 r_s^3-10962 r_c^4 r_s^4-1960 r_c^3 r_s^5+152 r_c^2 r_s^6+112 r_c r_s^7+11 r_s^8-\left(-3 r_c^2+2 r_c r_s+r_s^2\right)^2 \left(29 r_c^6+310 r_c^5 r_s+1067 r_c^4 r_s^2+692 r_c^3 r_s^3+179 r_c^2 r_s^4+22 r_c r_s^5+5 r_s^6\right),
\end{dmath*}
\begin{dmath*}
X^5 \C_{\chi_{c0} D_s} =1-25 r_c^2-18 r_c r_s-21 r_s^2+214 r_c^4+280 r_c^3 r_s+472 r_c^2 r_s^2+240 r_c r_s^3+138 r_s^4-2 \left(339 r_c^6+490 r_c^5 r_s+1183 r_c^4 r_s^2+1012 r_c^3 r_s^3+1033 r_c^2 r_s^4+402 r_c r_s^5+149 r_s^6\right)+441 r_c^8-888 r_c^7 r_s-536 r_c^6 r_s^2-2448 r_c^5 r_s^3+74 r_c^4 r_s^4+328 r_c^3 r_s^5+1808 r_c^2 r_s^6+960 r_c r_s^7+261 r_s^8-\left(9 r_c^5-27 r_c^4 r_s+2 r_c^3 r_s^2-14 r_c^2 r_s^3+21 r_c r_s^4+9 r_s^5\right)^2,
\end{dmath*}
\begin{dmath*}
X^5 \C_{\chi_{c0} D_s^*} =1+315 r_c^2+606 r_c r_s+303 r_s^2-2683 r_c^4-5808 r_c^3 r_s-4706 r_c^2 r_s^2-2336 r_c r_s^3-851 r_s^4+5293 r_c^6+11238 r_c^5 r_s+7227 r_c^4 r_s^2+5524 r_c^3 r_s^3+5379 r_c^2 r_s^4+2502 r_c r_s^5+725 r_s^6-2 \left(709 r_c^8+3706 r_c^7 r_s-2268 r_c^6 r_s^2-2050 r_c^5 r_s^3+738 r_c^4 r_s^4+502 r_c^3 r_s^5+444 r_c^2 r_s^6+210 r_c r_s^7+57 r_s^8\right)-4 \left(21 r_c^5-32 r_c^4 r_s+14 r_c^3 r_s^2+12 r_c^2 r_s^3-11 r_c r_s^4-4 r_s^5\right)^2,
\end{dmath*}
\begin{dmath*}
X^5 \C_{\chi_{c1} D_s} =1+16 r_c^2+30 r_c r_s+4 r_s^2-133 r_c^4-258 r_c^3 r_s-174 r_c^2 r_s^2-106 r_c r_s^3-17 r_s^4+237 r_c^6+504 r_c^5 r_s+47 r_c^4 r_s^2+32 r_c^3 r_s^3+195 r_c^2 r_s^4+120 r_c r_s^5+17 r_s^6+40 r_c^8-850 r_c^7 r_s-208 r_c^6 r_s^2-486 r_c^5 r_s^3-476 r_c^4 r_s^4+34 r_c^3 r_s^5-56 r_c^2 r_s^6-42 r_c r_s^7-4 r_s^8-(r_c-r_s)^4 \left(-15 r_c^3-5 r_c^2 r_s+3 r_c r_s^2+r_s^3\right)^2,
\end{dmath*}
\begin{dmath*}
X^5 \C_{\chi_{c1} D_s^*} =1-4 r_c^2+6 r_c r_s+8 r_s^2-69 r_c^4-72 r_c^3 r_s-62 r_c^2 r_s^2-24 r_c r_s^3-21 r_s^4+181 r_c^6-646 r_c^5 r_s-1133 r_c^4 r_s^2-612 r_c^3 r_s^3-93 r_c^2 r_s^4-6 r_c r_s^5+5 r_s^6+4 \left(96 r_c^8+499 r_c^7 r_s+236 r_c^6 r_s^2-495 r_c^5 r_s^3-340 r_c^4 r_s^4-35 r_c^3 r_s^5+20 r_c^2 r_s^6+15 r_c r_s^7+4 r_s^8\right)-\left(-3 r_c^2+2 r_c r_s+r_s^2\right)^2 \left(69 r_c^6+160 r_c^5 r_s+143 r_c^4 r_s^2-64 r_c^3 r_s^3-61 r_c^2 r_s^4+9 r_s^6\right),
\end{dmath*}
\begin{dmath*}
X^5 \C_{\chi_{c2} D_s} =1+2 \left(r_c^2-3 r_c r_s-3 r_s^2\right)-35 r_c^4+178 r_c^3 r_s+250 r_c^2 r_s^2+114 r_c r_s^3+21 r_s^4-837 r_c^6-2096 r_c^5 r_s-2819 r_c^4 r_s^2-2288 r_c^3 r_s^3-1187 r_c^2 r_s^4-336 r_c r_s^5-37 r_s^6+2 (3 r_c+r_s)^2 \left(143 r_c^6-297 r_c^5 r_s-251 r_c^4 r_s^2-22 r_c^3 r_s^3+133 r_c^2 r_s^4+87 r_c r_s^5+15 r_s^6\right)-(3 r_c+r_s)^4 \left(5 r_c^3-11 r_c^2 r_s+3 r_c r_s^2+3 r_s^3\right)^2,
\end{dmath*}
\begin{dmath*}
X^5 \C_{\chi_{c2} D_s^*} =1+6 \left(r_c^2+7 r_c r_s+5 r_s^2\right)-265 r_c^4-804 r_c^3 r_s-878 r_c^2 r_s^2-404 r_c r_s^3-89 r_s^4+1645 r_c^6+4182 r_c^5 r_s+5067 r_c^4 r_s^2+3652 r_c^3 r_s^3+1899 r_c^2 r_s^4+630 r_c r_s^5+77 r_s^6-4 \left(59 r_c^8-238 r_c^7 r_s-54 r_c^6 r_s^2+298 r_c^5 r_s^3+318 r_c^4 r_s^4+350 r_c^3 r_s^5+234 r_c^2 r_s^6+54 r_c r_s^7+3 r_s^8\right)-\left(-3 r_c^3-r_c^2 r_s+3 r_c r_s^2+r_s^3\right)^2 \left(199 r_c^4+58 r_c^3 r_s-18 r_c^2 r_s^2+10 r_c r_s^3+7 r_s^4\right).
\end{dmath*}
\end{dgroup*}
It is easy to see that in massless limit $r_{c,s}\to 0$ for all these coefficients we have $\C_{\Q\Ds}=1$.

\end{document}